\documentclass[aip,apl,reprint]{revtex4-1}

\usepackage{graphicx,dcolumn,bm,amsmath,amssymb,euscript}

\pdfoutput=1    

\newcommand{\VERSION}{30 June 2017}    

\usepackage[utf8]{inputenc}

\usepackage[colorlinks,citecolor=red,urlcolor=blue]{hyperref}

\graphicspath{{figures_pdf/submit/}}

\begin{document}

\title{Island dynamics and anisotropy during vapor phase epitaxy of m-plane {GaN}}

\author{Edith Perret}
\affiliation{Materials Science Division, Argonne National Laboratory, Argonne, IL 60439 USA}
\affiliation{University of Fribourg, Department of Physics and Fribourg Center for Nanomaterials, Chemin du Mus\'ee 3, CH-1700 Fribourg, Switzerland}
\author{Dongwei Xu}
\affiliation{Materials Science Division, Argonne National Laboratory, Argonne, IL 60439 USA}
\author{M. J. Highland}
\affiliation{Materials Science Division, Argonne National Laboratory, Argonne, IL 60439 USA}
\author{G. B. Stephenson}
\affiliation{Materials Science Division, Argonne National Laboratory, Argonne, IL 60439 USA}
\author{P. Zapol}
\affiliation{Materials Science Division, Argonne National Laboratory, Argonne, IL 60439 USA}
\author{P. H. Fuoss}
\altaffiliation[current address: ]{SLAC National Accelerator Laboratory, Menlo Park, CA 94025 USA}
\affiliation{Materials Science Division, Argonne National Laboratory, Argonne, IL 60439 USA}
\author{A. Munkholm}
\affiliation{Munkholm Consulting, Mountain View, CA 94043 USA}
\author{Carol Thompson}
\email[correspondence to: ]{cthompson@niu.edu}
\affiliation{Department of Physics, Northern Illinois University, DeKalb IL 60115 USA}

\date{\VERSION}

\begin{abstract}
Using \textit{in situ} grazing-incidence x-ray scattering, we have measured the diffuse scattering from islands that form during layer-by-layer growth of GaN by metal-organic vapor phase epitaxy on the $(1 0 \overline{1} 0)$ m-plane surface. The diffuse scattering is extended in the $(0 0 0 1)$ in-plane direction in reciprocal space, indicating a strong anisotropy with islands elongated along $[1 \overline{2} 1 0]$ and closely spaced along $[0 0 0 1]$. This is confirmed by atomic force microscopy of a quenched sample. Islands were characterized as a function of growth rate $G$ and temperature. The island spacing along $[0 0 0 1]$ observed during the growth of the first monolayer obeys a power-law dependence on growth rate $G^{-n}$, with an exponent $n = 0.25 \pm 0.02$. Results are in agreement with recent kinetic Monte Carlo simulations, indicating that elongated islands result from the dominant anisotropy in step edge energy and not from surface diffusion anisotropy. The observed power-law exponent can be explained using a simple steady-state model, which gives $n = 1/4$.
\end{abstract}

\maketitle

The competition among atomic-scale processes at the surface of a growing epitaxial film produces a fascinating array of growth mechanisms, morphologies and crystal growth modes.\cite{1993_Tsao_MatFundMBE,2010_Jackson_KinProc_2ndedition}
Study of these modes can not only reveal the nature of the critical processes but also allow rational design of methods to synthesize high quality films and heterostructures with the interface morphology, dopant distributions, and controlled defect levels needed for devices.\cite{2013_DenBaars_ActaMater61_945, 2012_Browne_JVacSciTechnolA30_041513}
Depending upon the balance between rates of deposition, surface diffusion on terraces, attachment at steps, and nucleation of islands on terraces, the growth mode can vary among step-flow, layer-by-layer, and three-dimensional.\cite{1993_Tsao_MatFundMBE}
Because the bonding geometries, energies, and diffusion barriers are all typically strong functions of crystal surface orientation, growth modes also vary with orientation.\cite{2014_Perret_APL105_051602} 
During layer-by-layer growth, in which islands nucleate and coalesce to form each molecular layer of the crystal in succession, observation of the oscillatory surface morphology produces an especially sensitive measure of this balance of surface processes.

\begin{figure}
\includegraphics[width=0.9\linewidth]{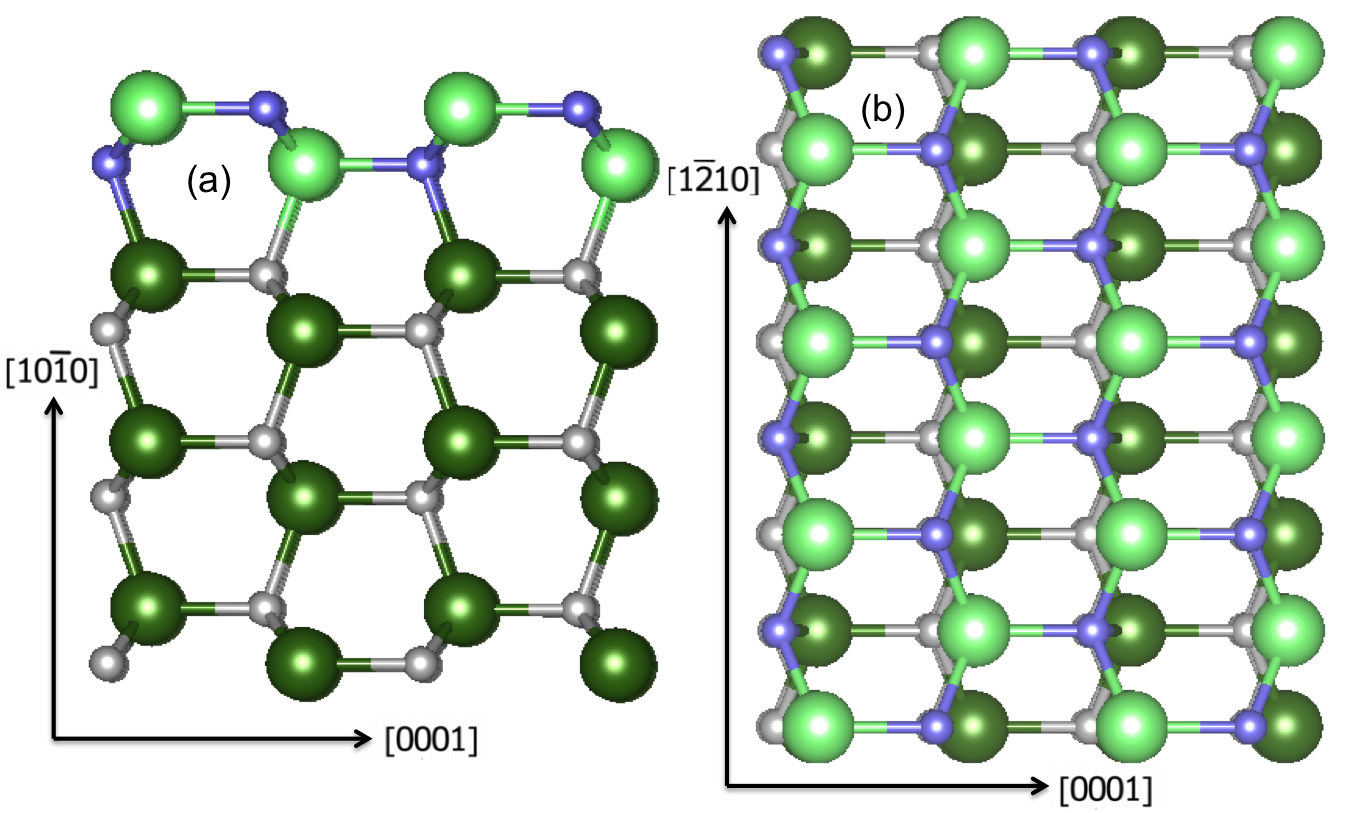}
\caption{\label{fig:planes}Structure of an ideally terminated GaN m-plane $(1 0 \overline{1} 0)$ surface. (a) Cross-section with m-plane surface on top, showing corrugations of surface. (b) Plan view of m-plane surface. Top layer Ga and N atoms are light green and blue, respectively; lower layer atoms are darker.
}
\end{figure}

Because of their potential importance in improving the performance of optoelectronic devices, growth of GaN films in non-polar and semi-polar orientations has received increasing attention.\cite{2013_DenBaars_ActaMater61_945}
These non-basal-plane orientations of the wurtzite structure have in-plane  surface anisotropy, often resulting in complex growth behavior and surface morphologies.
In particular, the m-plane $(1 0 \overline{1} 0)$ surface of GaN has been the subject of fundamental study.
Its ideal structure is shown in Fig.~\ref{fig:planes}.
First-principles-based theory for this surface in vacuum predicts relaxations from the ideal structure,\cite{1996_Jaffe_PRB53_R4209} 
as well as highly anisotropic activation barriers for surface diffusion.\cite{2009_Lymperakis_PRB79_241308,2010_Jindal_JAP107_054907}
The nature of GaN surfaces in the metal-organic vapor phase epitaxy (MOVPE) environment studied here is affected by attachment of NH$_x$ species.\cite{2012_Walkosz_PRB85_033308}
A recent comparison\cite{2017_Xu_JCP146_144702} of kinetic Monte Carlo (KMC) simulations and experiments on MOVPE of m-plane GaN indicates that  anisotropy in step edge energies rather than anisotropy in diffusion barriers dominates surface morphology under typical MOVPE conditions.
Here we present results of an {\it in-situ} surface x-ray scattering study of the island shape and nucleation density during MOVPE of GaN on the m-plane surface as a function of temperature $T$ and growth rate $G$.
\textit{In-situ} x-ray scattering and \textit{ex-situ} atomic force microscopy both show islands elongated along  the $[1 \overline{2} 1 0]$ direction, with an anisotropy that increases as growth rate decreases.
The island spacing $S_z$ in the $[0 0 0 1]$ direction has a very weak power law dependence upon growth rate $S_z \propto G^{-n}$ with an exponent of $n = 0.25 \pm 0.02$, in agreement with simulation results.\cite{2017_Xu_JCP146_144702}
We present a simple steady-state analysis to explain this exponent.

\begin{figure}
\includegraphics[width=0.95\linewidth]{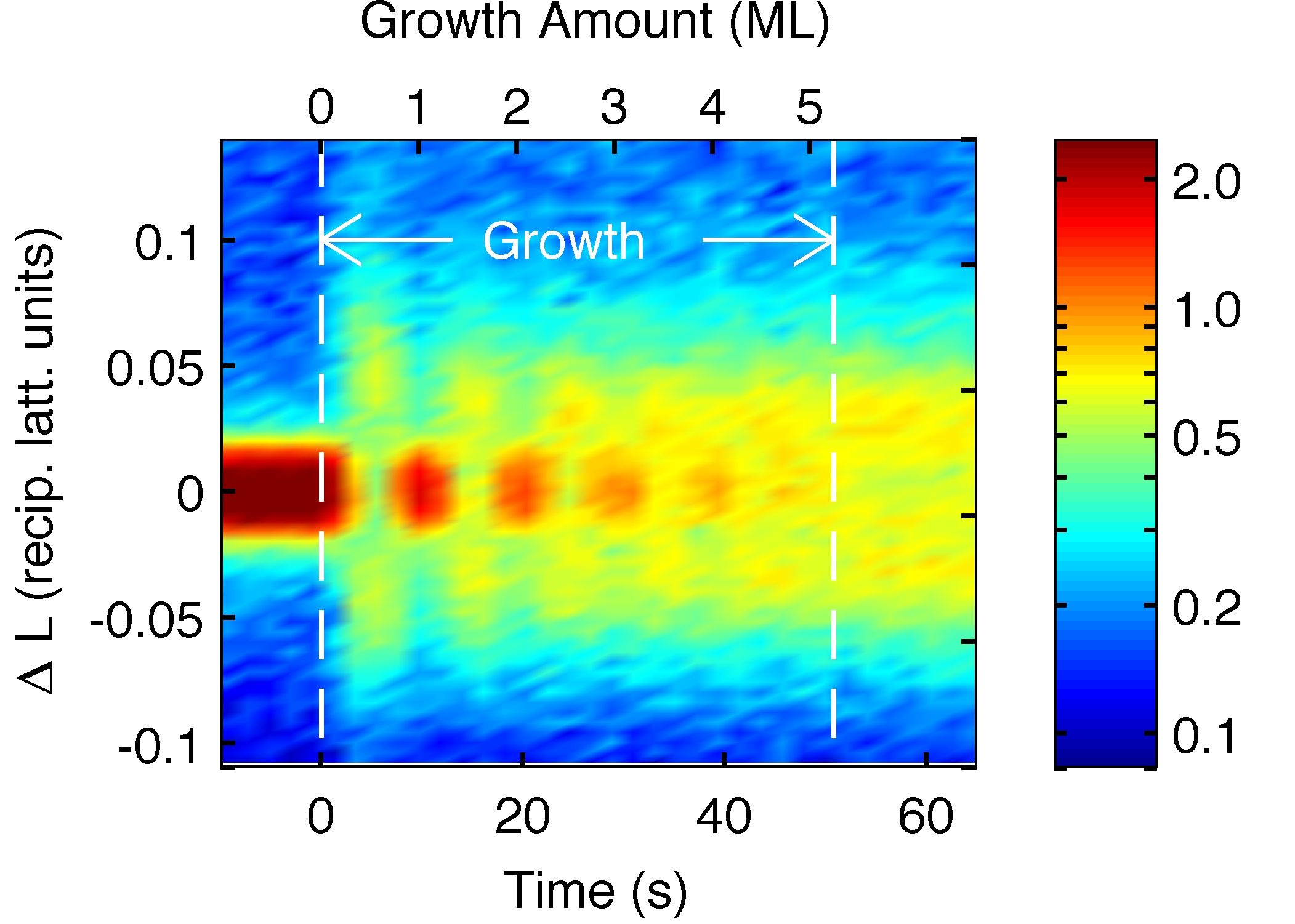}
\caption{\label{fig:pcolor}Typical x-ray intensity distribution in the $L$ direction around the $(H \, 0 \, \overline{H} \, \overline{2})$ CTR near $H = 0.5$, as a function of time before, during, and after 50 s of layer-by-layer growth at $T = 893$~K and $G = 0.31$~\AA/s. Satellite peaks of diffuse intensity appear around the CTR at half-monolayer coverages, reflecting the correlated island spacings. 
}
\end{figure}

\begin{figure}
\includegraphics[width=0.85\linewidth]{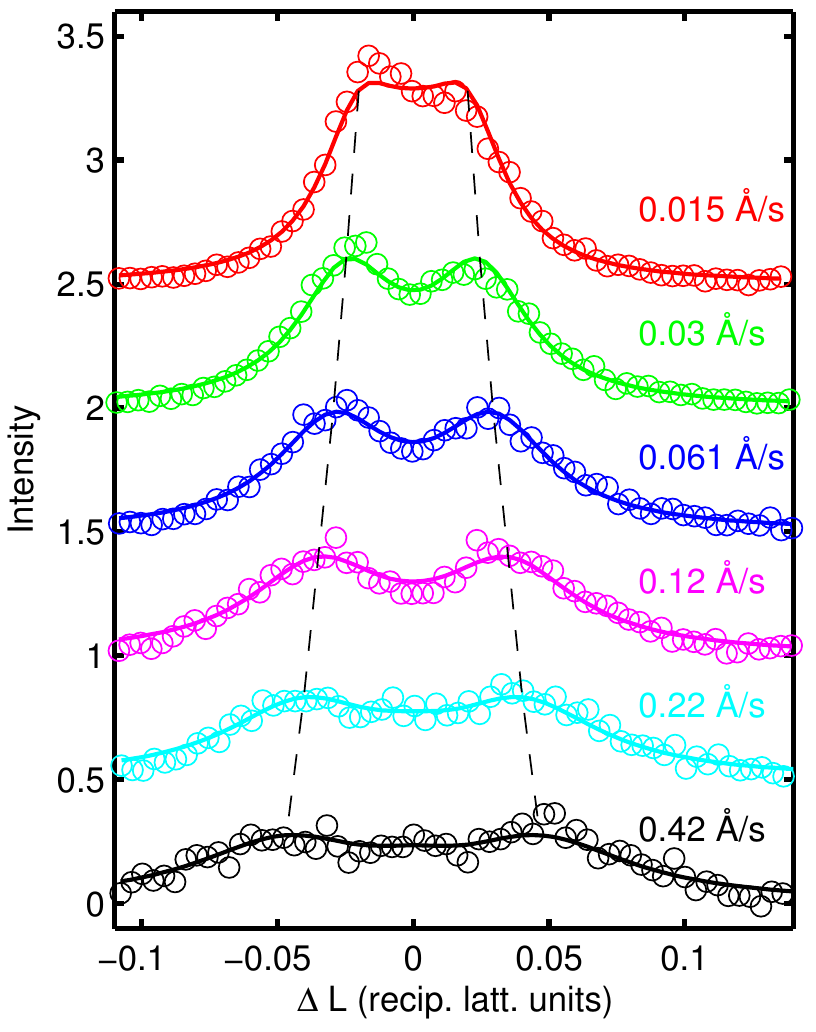}
\caption{\label{fig:fits}Typical diffuse intensity distributions around the CTR in the $L$ direction at 0.5 ML of growth, for $T = 893$~K at the indicated growth rates, showing the variation in the satellite peak positions with growth rate. Also shown are fits to extract the peak positions $\Delta L_{pk}$. Values of $\Delta L_{pk}$ are indicated by dashed lines. Curves are offset for clarity.}
\end{figure}

\begin{figure*}
\includegraphics{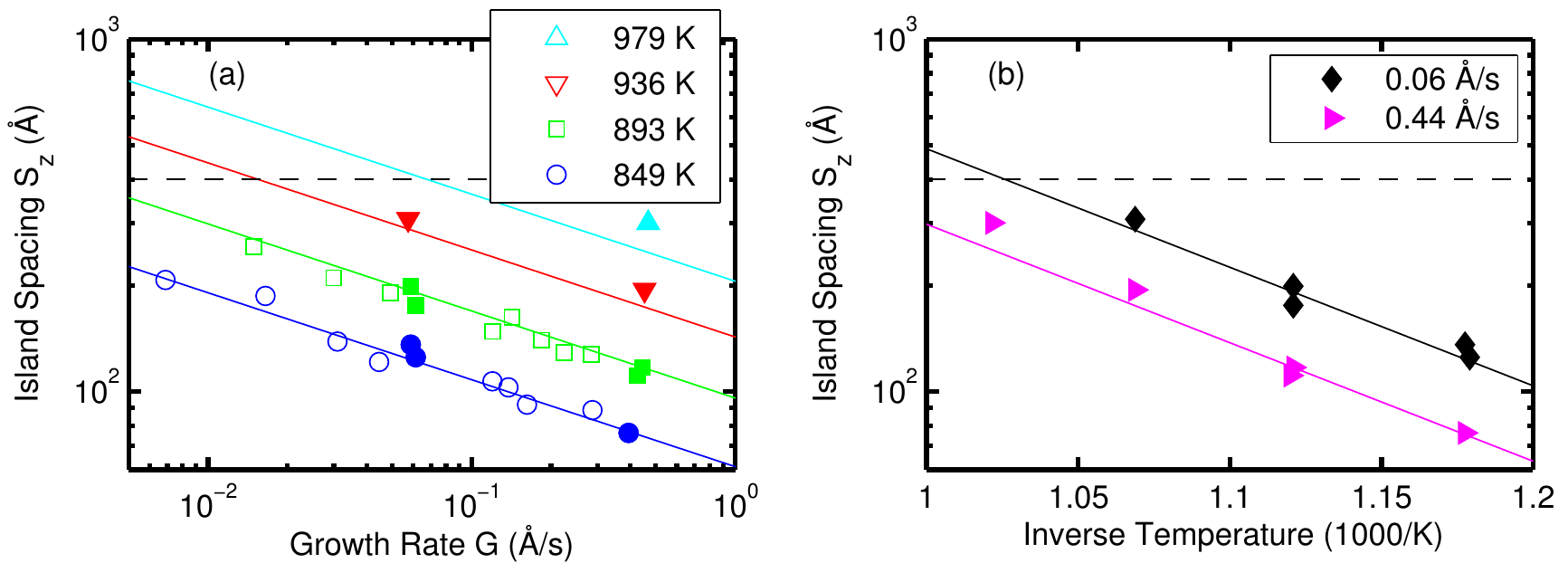}
\caption{\label{fig:spacing}(a) Island spacing at 0.5 ML vs. growth rate at fixed temperature, showing the power law dependence. (b) Island spacing at 0.5 ML vs. inverse temperature for selected growth rates (solid symbols in (a)), showing the Arrhenius dependence. Island spacings were limited to values less than the terrace width $W = 400$~\AA, dashed line, because of the transition to step-flow growth. Lines show a fit of all points to Eq.~(1) to extract values of $n$, $E_S$, and $G_S$.}
\end{figure*}

We used real-time grazing incidence x-ray scattering experiments to characterize the surface structures that form during homoepitaxy of GaN by MOVPE.
The substrate was a GaN single crystal with a surface oriented 0.4 degrees from the $(1 0 \overline{1} 0)$ planes,
giving terraces of width $W = 400$~\AA~ along the [0 0 0 1] direction separated by single-monolayer (ML) steps of height $d_{10\bar{1}0} = \sqrt{3} a_0/2 = 2.76$~\AA.
Experimental methods were the same as described in a previous study.\cite{2014_Perret_APL105_051602,Supplemental}
Triethylgallium (TEGa) and ammonia (NH$_3$) were used as precursors and nitrogen as carrier gas.
Growth rate was controlled by varying the supply of TEGa, with a large excess of NH$_3$.
Substrate temperature was determined within $\pm 5$~K by calibration using thermal expansion of a standard sapphire substrate measured by optical interferometry.\cite{2017_Ju_RSI88_035113} 
In the previous study,\cite{2014_Perret_APL105_051602} we found conditions under which layer-by-layer growth occurs by observing the extent to which the intensity of the crystal truncation rod (CTR) scattering oscillates in time during growth, with maxima at the completion of each ML of growth.
Here we study the diffuse scattering that occurs around the CTR when islands are present on the surface between the completion of each ML,
allowing us to determine the spacing of the islands and how it varies with growth conditions.
Figure \ref{fig:pcolor} shows typical x-ray diffuse scattering from islands nucleating and coalescing during MOVPE on m-plane GaN in the layer-by-layer growth regime.
Intensity is plotted as a function of time $t$ and distance $\Delta L$ from the CTR in the in-plane $L$ direction.\cite{Supplemental}
As found previously,\cite{2014_Perret_APL105_051602} the CTR peak at $\Delta L = 0$ in Fig.~\ref{fig:pcolor} oscillates strongly in time after growth is initiated at $t = 0$. 
When the intensity of the CTR is at a minimum, corresponding to half-filled layers, we observe diffuse scattering extending in the in-plane $L$ direction around the CTR, which originates from islands on the surface.
The diffuse scattering shows peaks on each side of the CTR, indicating highly correlated island positions.
We studied many growth conditions using the same sample by growing only a few monolayers under each condition, and then annealing the sample at 1230~K for 180~s to recover the surface to its initial state for further growth studies.

Figure \ref{fig:fits} shows the distribution of this diffuse intensity in $L$ at the first minimum in the CTR intensity, i.e. after 0.5 ML of growth, for various growth rates at $T = 893$~K.
Average island spacings were extracted by fitting the data to obtain the positions $\pm \Delta L_{pk}$ of the satellite peaks in the diffuse scattering (dashed lines in Fig.~\ref{fig:fits}). 
The displacement $\Delta L_{pk}$ of the peaks from the CTR is inversely proportional to the average island spacing, $S_z = c_0/\Delta L_{pk}$, in the $[0 0 0 1]$ direction, with $c_0 = 5.18$~\AA.
Typical fits of the diffuse scattering are shown in Fig.~\ref{fig:fits}. 
Details of the fitting method are given in Supplemental Material.\cite{Supplemental} 

The average island spacing $S_z$ at 0.5 ML coverage is plotted versus growth rate $G$ at fixed temperature $T$ in Fig.~\ref{fig:spacing}(a), and versus inverse $T$ at fixed $G$ in Fig.~\ref{fig:spacing}(b). 
Note that the average island spacing is smallest for growth of the first layer and increases for subsequent layers, as indicated by the decrease in $\Delta L_{pk}$ apparent in Fig.~\ref{fig:pcolor}. This behavior is typical of that seen in other surface scattering studies of layer-by-layer growth.\cite{1996_Kisker_JCrystGrowth163_54,2009_Ferguson_PRL103_256103}

The island spacings can be fit to a power-law dependence on $G$ and an Arrhenius dependence on $T$, expressed by
\begin{equation}
S_z/a_0 = (G/G_S)^{-n} \exp(-nE_S/kT).
\label{eq:S_z}
\end{equation}
Here we have scaled $S_z$ by the lattice parameter $a_0 = 3.19$~\AA~ rather than $c_0$ to facilitate comparison with the KMC study.\cite{2017_Xu_JCP146_144702}
A similar expression is obtained from analysis of island nucleation spacings.\cite{2006_Evans_SurfSciRep61_1}
The three parameter values obtained from a fit to all 26 island spacings at 0.5 ML in the layer-by-layer regime shown in Fig.~\ref{fig:spacing}(a) are: an exponent $n = 0.25 \pm 0.02$, an activation energy $E_S = 2.70 \pm 0.18$~eV, and a growth rate scale factor $\log_{10}[G_S $(\AA/s)$] = 21.2 \pm 1.0$. 
The dependence of $S_z$ on $G$ and $T$ is consistent with the boundary between step-flow and layer-by-layer growth modes determined from the amplitude of CTR oscillations in the previous study,\cite{2014_Perret_APL105_051602} assuming that the boundary corresponds to an average island spacing at 0.5 ML equal to the average terrace width, $S_z = W = 400$~\AA.\cite{Supplemental}

\begin{figure}
\includegraphics[width=0.9\linewidth]{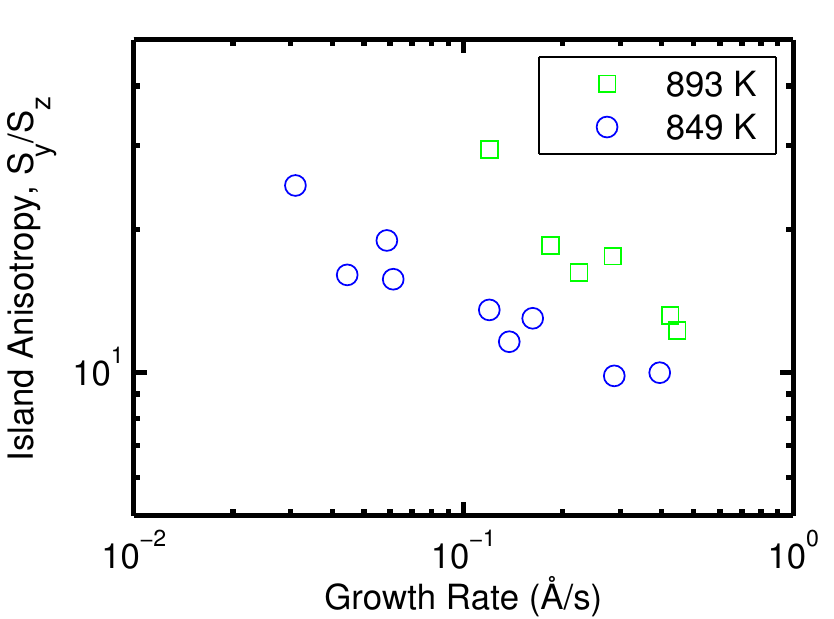}
\caption{\label{fig:aspect_ratio}Ratio of island spacing $S_y$ along $[1 \overline{2} 1 0]$ to island spacing $S_z$ along $[0001]$, as a function of growth rate, for two temperatures. Anisotropy increases as growth rate decreases.
}
\end{figure}

A strong anisotropy in the island shape and spacing is apparent in the diffuse scattering, which is extended in reciprocal space only in the $(0 0 0 1)$ in-plane direction. 
In the perpendicular $(1 \overline{2} 1 0)$ in-plane direction, the scattering is peaked at the CTR position.
For higher growth rates and lower temperatures, the width of this peak increases measurably above the lower limit imposed by the experimental resolution, reflecting a decreasing island spacing $S_y$ along $[1 \overline{2} 1 0]$. 
For these conditions we can extract\cite{Supplemental} an approximate value of $S_y$ from the peak width.
Figure \ref{fig:aspect_ratio} shows the island spacing anisotropy $S_y/S_z$ as a function of growth rate for two temperatures.
The higher anisotropy observed at lower growth rates, i.e. closer to equilibrium, indicates that anisotropy is an equilibrium rather than kinetically driven phenomenon.

\begin{figure}
\includegraphics[width=0.9\linewidth]{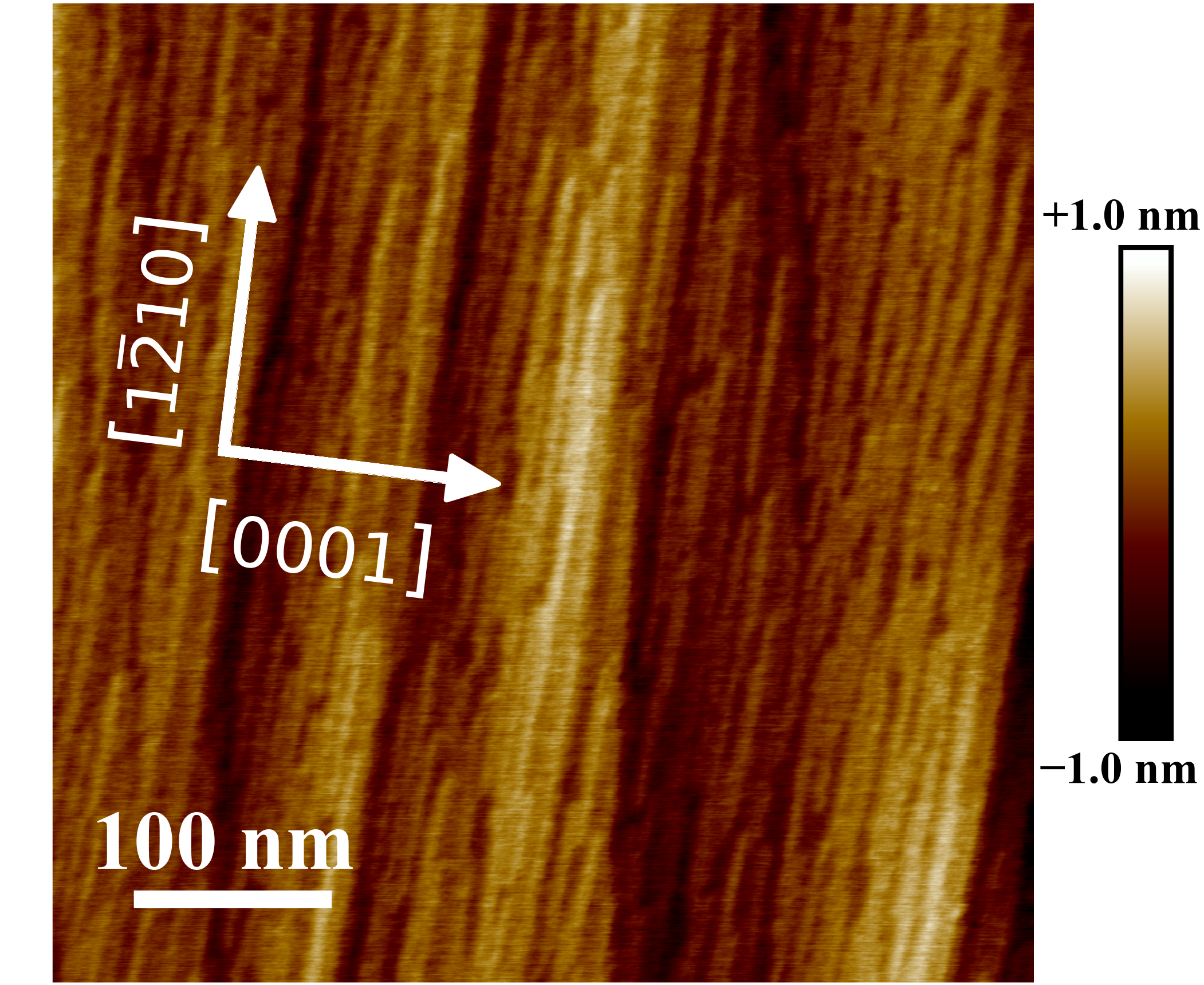}
\caption{\label{fig:AFM_islands}Atomic-force microscopy image of an m-plane surface quenched after 0.5 ML of growth, showing terraces covered by narrow islands with spacing $S_z \approx 14$~nm in the $[0 0 0 1]$ direction.}
\end{figure}

To image the island anisotropy, we grew 0.5 ML under layer-by-layer conditions at a temperature sufficiently low ($T = 849$~K, $G = 0.06$~\AA/s) that the sample could be quenched to room temperature after growth without any further change in the island structure, as monitored by the diffuse scattering. 
Figure \ref{fig:AFM_islands} shows an atomic force microscopy image of the islands on the 0.5-ML quenched surface. 
The strong elongation of the islands along $[1 \overline{2} 1 0]$ is apparent,
and the observed island spacing along $[0 0 0 1]$ agrees with the diffusion scattering determination.
 
Our experimental results are in good agreement with recent KMC simulations of MOVPE growth on m-plane GaN.\cite{2017_Xu_JCP146_144702} In the KMC study, islands with similar anisotropy (elongated perpendicular to $[0 0 0 1]$) were observed during layer-by-layer growth, and the island spacing $S_z$ had the same dependence on $G$ and $T$ given by Eq.~(1). The island spacing power-law exponent was found to be $n = 0.24 \pm 0.01$, in agreement with that observed here. The higher island spacing anisotropy as equilibrium is approached (lower growth rates) shown in Fig.~\ref{fig:aspect_ratio} is consistent with the conclusion of the KMC study\cite{2017_Xu_JCP146_144702} that island shape is determined primarily by anisotropy in equilibrium step edge energy, rather than by anisotropy in surface diffusivity.

The $n = 0.25$ power-law exponent for the island spacing $S_z$ dependence on growth rate $G$ can be understood by considering the value of $S_z$ needed to balance the rate of attachment of adatoms to existing islands and the rate of adatom deposition. If we assume that all characteristic lengths of the island structure scale with $S_z$, and that surface transport occurs by diffusion rather than evaporation/condensation (as is the case in both the experiments and simulations), the rate of adatoms diffusing to existing island edges is proportional to $D / S_z^2$, where $D$ is the adatom diffusivity. The rate of adatom deposition onto terraces between islands is proportional to $J S_z^2$, where $J = G / V_0$ is the deposition flux per unit area and $V_0 = (\sqrt{3}/4) a_0^2 c_0$ is the molecular volume. When these two rates are balanced, the steady state island spacing is thus
\begin{equation}
    S_z \propto \left (\frac{D V_0}{G} \right ) ^{1/4}.
    \label{eq:S_ss}
\end{equation}
This result can be obtained for anisotropic or isotropic island structures and diffusivities.\cite{unpublished}
This gives an $n = 1/4$ power-law dependence on $1/G$, which agrees with the value of $n$ observed in our experiments and recent simulations\cite{2017_Xu_JCP146_144702} for $S_z$ at 0.5 ML, expressed by Eq.~(\ref{eq:S_z}).
The values we observe for $E_S$ and $G_S$ in Eq.~(\ref{eq:S_z}) are also in reasonable agreement with a more quantitative version of Eq.~(\ref{eq:S_ss}).\cite{unpublished}

In summary, we found that islands formed during layer-by-layer growth of GaN by MOVPE on the $(1 0 \overline{1} 0)$ m-plane surface are elongated perpendicular to $[0 0 0 1]$. The island spacing along $[0 0 0 1]$ obeys a power-law dependence on growth rate $G^{-n}$, with an exponent $n = 0.25$ consistent with simulations\cite{2017_Xu_JCP146_144702} and with a simple steady-state analysis. The very weak dependence of island spacing on $G$ indicates that island spacing can be most effectively controlled by changing growth temperature. Because island shape is controlled by step edge energy, the surfactant behavior of dopants such as Si may have a large effect on surface morphology.\cite{2000_Munkholm_APL77_1626}

\begin{acknowledgments}
Work supported by the U.S. Department of Energy (DOE), Office of Science, Office of Basic Energy Sciences, Division of Materials Sciences and Engineering. Use of beamline 12ID-D of the Advanced Photon Source, a DOE Office of Science User Facility operated for the Office of Science by Argonne National Laboratory, was supported under contract DE-AC02-06CH11357.
\end{acknowledgments}

\bibliography{bibliography/2017_Perret_GaN_f}

\end{document}